# Data-integration with pseudoweights and survey-calibration: application to developing US-representative lung cancer risk models for use in screening

Lingxiao Wang[1,2*], Yan Li[2], Barry I. Graubard[1,^], Hormuzd A. Katki[1,^]

[1]National Cancer Institute, Division of Cancer Epidemiology & Genetics, Biostatistics Branch, U.S.A.
[2]Department of Statistics, University of Virginia, U.S.A.
[3]The Joint Program in Survey Methodology and Department of Epidemiology and Biostatistics, University of Maryland, College Park, U.S.A.
*Correspondence: Lingxiao Wang, Department of Statistics, University of Virginia, Charlottesville, VA 22904. Email: lingxiao.wang@virginia.edu
^Two authors contributed equally

**ABSTRACT**

Accurate cancer risk estimation is crucial to clinical decision-making, such as identifying high-risk people for screening. However, most existing cancer risk models incorporate data from epidemiologic studies, which usually cannot represent the target population. While population-based health surveys are ideal for making inference to the target population, they typically do not collect time-to-cancer incidence data. Instead, time-to-cancer specific mortality is often readily available on surveys via linkage to vital statistics. We develop calibrated pseudoweighting methods that integrate individual-level data from a cohort and a survey, and summary statistics of cancer incidence from national cancer registries. By leveraging individual-level cancer mortality data in the survey, the proposed methods impute time-to-cancer incidence for survey sample individuals and use survey calibration with auxiliary variables of influence functions generated from Cox regression to improve robustness and efficiency of the inverse-propensity pseudoweighting method in estimating pure risks. We develop a lung cancer incidence pure risk model from the Prostate, Lung, Colorectal, and Ovarian (PLCO) Cancer Screening Trial using our proposed methods by integrating data from the National Health Interview Survey (NHIS) and cancer registries.

# 1 INTRODUCTION

The US Preventive Services Task Force (USPSTF) recommends lung cancer screening for ever-smokers ages 50-80, have smoked at least 20 pack-years, and have not quit smoking for more than 15 years (Krist et al., 2021). However, numerous studies suggest that use of individualized risk models would identify many more ever-smokers with high benefit, making screening more effective and efficient while reducing disparities because high-benefit African Americans are disproportionately ineligible for screening by the USPSTF criteria (Katki et al., 2016; Landy et al., 2020). Consequently, individualized risk models have been recommended for use in lung screening by many US medical societies (Mazzone et al., 2018), the UK (Oudkerk et al., 2017), and Canada (Tammemägi et al., 2021). However, many existing risk models have been shown to be inaccurate in US cohorts, especially for minorities (Katki et al., 2018). This could be because all recommended lung cancer risk models were fit to data from epidemiologic cohorts that recruited volunteers who do not represent the US population. Lack of representativeness is a serious problem due to "healthy volunteer effects", where disease incidence rates in cohorts can be substantially lower than that in the target population (Pinsky et al., 2007). As a result, risk models fit to the cohort may not be generalizable to the target population and therefore yield biased risk estimation.

A population representative pure risk model of lung cancer incidence is a key component of unbiased individualized risk estimation. Several methods have been proposed to improve generalizability of cohort pure risk models by using vital statistics of cancer incidence from external data sources (Gail, 1989; Zheng et al., 2022), but assume common hazard ratios (HRs) of cancer incidence between the observed cohort and the target population. However, we found that HRs for mortality were not generalizable from the cohort to the target population (Wang et al., 2022), raising a need for methods that do not make strong HR generalizability assumptions.

Our research is driven by needing to improve the generalizability of pure risk models for lung cancer incidence from large-scale, non-representative cohorts without a strong assumption regarding the equality of HRs between the cohort and the target population. The Prostate, Lung, Colorectal, and Ovarian (PLCO) Cancer Screening Trial provides one of the largest cohorts for lung cancer research in the US. PLCO recruited 149,905 volunteered participants ages 55-74 years at 10 centers across the U.S from 1993-2001 and followed participants for cancer and mortality up to 2015. However, PLCO substantially underrepresented racial/ethnic minorities and less educated people compared to the target US population within the same age range (Pinsky et al., 2007). The number of lung cancer cases in PLCO for minorities is small. Moreover, the averaged lung cancer risks estimated from the naïve cohort risk model within racial/ethnic groups may not match with the observed lung cancer incidence rates in the US (details in Section 5).

In contrast to volunteer-based cohort studies, cross-sectional population-based health surveys randomly select samples from the target population ensuring that the resulting weighted samples are representative of the target population. The inference can provide design consistent results with appropriate incorporation of the sample design in the estimation. Furthermore, national health surveys usually oversample minority groups to improve efficiency of the subpopulation analyses. Also, national health surveys usually have most risk-factors and self-reported disease status. Recognizing the strengths of both volunteer samples (such as cohorts) and survey samples, extensive methodological research has been conducted to enhance the representativeness of volunteer samples by "pseudoweighting" methods that use a representative survey sample as the reference, such as the Inverse Propensity Weighting (IPW) methods (Chen et al., 2020; Elliott and Valliant, 2017; Wang et al., 2021). Aiming to further reduce the bias of the IPW method due to propensity model misspecification and improve the efficiency, Elliott (2009) and Robbins et al.

(2021) combine the pseudoweighted volunteer sample with the weighted survey sample for cross-sectional regression analyses when both outcome and predictors are available in the two samples.

A lung cancer incidence pure risk model cannot be fit to the combined (pseudo)weighted sample due to the ubiquitous practical problem that population-based health surveys generally do not follow participants for time-to-disease incidence outcomes. Instead, population-based national health surveys, such as US health surveys (National Center for Health Statistics, 2009) and the Health Survey for England (Mindell et al., 2012), are generally linked to national vital statistics to obtain time-to-cancer specific mortality outcomes for eligible survey sample individuals, which can be strongly associated with time-to-cancer incidence outcomes, especially for highly fatal cancers, such as lung cancer. This motivates us to improve the accuracy and efficiency of the IPW pseudoweighting lung cancer incidence risk estimation by integrating individual-level data on lung cancer mortality and predictors that are available in both the cohort and the survey samples.

In this paper, we develop novel calibrated pseudoweighting methods for cancer incidence pure risk estimation that (1) impute missing cancer incidence times for survey sample individuals by using cancer mortality outcomes, and (2) improve robustness as well as efficiency of the IPW pseudoweighted risk estimators by calibrating the pseudoweighted cohort to the combined (pseudo)weighted cohort and survey sample on novel calibration auxiliary variables generated from influence functions (Breslow et al., 2009). Different from the existing methods that synthesize the external summary information of the outcome (or a proxy) with the cohorts to improve the accuracy and efficiency of risk estimation (Zheng et al., 2022; 2023; Wang et al., 2022), or the standard survey calibration methods that borrow strength from summary statistics of covariates in the population (Deville and Särndal, 1992; Valliant et al., 2013), the proposed calibrated pseudoweighting methods leverage individual-level data on cancer mortality and predictors from the survey to generate influence functions as the auxiliary variables for cohort calibration. Calibration on influence functions is a powerful and general approach to improving efficiency without inducing bias in sample weighted regression coefficient estimation, such as hazard ratios (HR's) (Breslow and Hu, 2018; Han, 2016; Lumley et al., 2011; Lumley, 2018, Shin et al., 2020) under two-phase sampling designs in epidemiologic studies. In finite population inference, we show that our calibrated pseudoweighting approaches are less sensitive to propensity model misspecification and more efficient than the standard IPW methods. The proposed calibrated pseudoweighting methods can also be combined with other exiting methods that borrow vital statistics of cancer incidence from the population.

Our approach provides robust and efficient pure risk estimation for lung cancer by integrating the volunteer based PLCO cohort with both incidence and mortality outcomes, the US National Health Interview Survey (NHIS) with mortality outcomes but not incidence outcomes (which is typical), and lung cancer data from the US vital statistics. We are unaware of any existing method that leverages individual-level cancer mortality outcomes to improve risk estimation for cancer incidence.

Table 1 Data structure of the cohort, survey sample, and population registry

| Data Source | Sample weight | Lung Cancer Incidence(Outcome) | Lung Cancer Mortality | Predictors Demographics | Other predictors |
|---|---|---|---|---|---|
| PLCO (Cohort) | x | √ | √ | √ | √ |
| NHIS (Survey) | √ | x | √ | √ | √ |
| Vital statistics | 1 | √ | √ | √ | x |

The paper is outlined as follows. We first describe the framework of survival model in the finite target population, and cohort and survey sample notation and then introduce the proposed calibrated pseudoweighting methods and Jackknife variance estimation. Simulations investigate the robustness and efficiency of the proposed methods compared with the existing IPW methods under correctly specified and misspecified propensity models. We develop a risk model to estimate the pure risk of lung cancer incidence for people ages 55-74 in the U.S. combining cohort data from PLCO, NHIS 1997-2000, and the lung cancer data form the US vital statistics using our proposed calibration methods. Finally, we conclude the paper with a few remarks.

## 2  Notation

Let $FP = \{1, 2, \cdots, M\}$ be the set of $M$ individuals for the target finite population ($FP$). We are interested in estimating pure risk (Gail, 2011; Shin et al., 2020; Zheng et al., 2022) for an individual in the $FP$, i.e., the probability of cancer incidence by time $t$, $r^{FP}(t) = P(T \leq t \mid FP)$ where $T$ is the time to cancer incidence of interest. Suppose each individual $i \in FP$ has a vector of non-time dependent covariates $\mathbf{z}_i$, a potential cancer incidence time $T_i$ or a censoring time $C_i$ that is independent from $T_i$ given $\mathbf{z}_i$ (e.g., drop-off due to other causes of mortality). The observed time $X_i = \min(T_i, C_i)$. Denote the counting process of disease incidence $N_i(t)$, ($N_i(t) = 1$ if $T_i < t$, and $T_i = X_i$; and $N_i(t) = 0$ otherwise), and the at-risk process, $Y_i(t)$. For the time-on-study metric, $Y_i(t) = 1$ if $X_i \geq t$; and $Y_i(t) = 0$ otherwise. The time metric is follow-up time with a maximum of $C_0$ (i.e., administrative censoring time). The disease status for each individual $i$ during a given follow-up period is $D_i$ ($= 1$ if $T_i = \min(X_i, C_0)$, and 0 otherwise), with $C_0$ being the maximum of the follow-up time (i.e., administrative censoring time). Under a Cox regression model (Cox 1972), we have

$$r^{FP}(t, \mathbf{z}, \boldsymbol{\beta}^{FP}, \Lambda_0^{FP}) = P(T \leq t \mid, \mathbf{z}, FP) \\ = 1 - \exp\{-\Lambda_0^{FP}(t) \cdot \exp(\mathbf{z}^T \boldsymbol{\beta}^{FP})\}, \quad (2.1)$$

where $\boldsymbol{\beta}^{FP}$ is a vector of $FP$ log transformed hazard-ratios (log-HR's) under a Cox regression model in the $FP$, which obtained by solving the $FP$ estimating equation for $\boldsymbol{\beta}$

$$\boldsymbol{U}^{FP}(\boldsymbol{\beta}) = M^{-1} \sum_{i \in FP} \int_0^\infty \{\mathbf{z}_i - S^{(0)}(\tau, \boldsymbol{\beta})^{-1} S^{(1)}(\tau, \boldsymbol{\beta})\} dN_i(\tau) = \mathbf{0}, \quad (2.2)$$

where $S^{(u)}(\tau, \boldsymbol{\beta}) = M^{-1} \sum_{i \in FP} Y_i(\tau) \exp\{\boldsymbol{\beta}^T \mathbf{z}_i\} \cdot \mathbf{z}_j^{\otimes u}$, with $u = 0, 1$, and $d\widetilde{N}_i(\tau) = \widetilde{N}_i(\tau) - \widetilde{N}_i(\tau-)$ is the increment of $\widetilde{N}_i$ for the disease-specific mortality at time $\tau$, with $\widetilde{N}_i(\tau-) = \lim_{\epsilon \to 0^+} \widetilde{N}_i(\tau - \epsilon)$, and $\Lambda_0^{FP}(t)$ is the cumulative baseline hazard in the $FP$ is

$$\Lambda_0^{FP}(t) = \int_0^t \frac{\sum_{i \in FP} dN_i(\tau)}{\sum_{i \in FP} Y_i(\tau) \exp(\mathbf{z}_i^T \boldsymbol{\beta}^{FP})} \tag{2.3}$$

Denote $\widetilde{T}_i$ as the death time due to the disease of interest for individual $i \in FP$. The mortality status in the follow-up period $(0, C_0)$ is $\widetilde{D}_i$, ($\widetilde{D}_i = 1$ if $\widetilde{T}_i = \min(\widetilde{X}_i, C_0)$, and $\widetilde{D}_i = 0$ otherwise, where $\widetilde{X}_i = \min(\widetilde{T}_i, C_i)$). The actual and observed gap time between the disease incidence of interest and the disease-specific mortality is $\Delta T_i = \widetilde{T}_i - T_i$ and $\Delta X_i = \widetilde{X}_i - X_i$ respectively. Similar to disease incidence, we denote the counting process and the at-risk process of disease mortality $\widetilde{N}_i(t)$, and $\widetilde{Y}_i(t)$, respectively.

Suppose in the $FP$, we do not observe individual-level data $(X_i, D_i, \widetilde{X}_i, \widetilde{D}_i, \mathbf{z}_i, i \in FP)$, and therefore cannot obtain $\widehat{\boldsymbol{\beta}}^{FP}$, $\Lambda_0^{FP}(t)$ or $r^{FP}(t)$. We only have the total number of individuals $M$, and summary statistics of the disease (e.g., vital statistics of lung cancer), such as number of cases by the follow-up time $t$ in mutually exclusive subpopulations defined by demographic variables (i.e., age/sex/race), $M_{1,g}(t) = \sum_{i \in FP_g} N_i(t)$, with $FP_g$, $g = 1 \cdots, G$ being the $g^{\text{th}}$ subpopulation, and similarly, the disease-specific, and all-cause mortality rates in the subpopulations by the follow-up time.

Let $s_c \subset FP$ denote a cohort with $n_c$ individuals. We define a random indicator variable $\delta_i^{(c)}$ ($= 1$ if $i \in s_c$; 0 otherwise) that specifies which individuals in $FP$ participate in $s_c$. Note that $FP$ and $s_c$ also denote sets of indices for the $FP$ and the cohort, respectively. We assume there is a positive underlying cohort participation rate for each $i \in s_c$, $\pi_i^{(c)} \equiv P(i \in s_c \mid FP) =$ $E_c\left(\delta_i^{(c)} \mid FP\right) > 0$, where the expectation $E_c$ is with respect to the unknown random cohort sample participation process from $FP$ (Wang et al. 2021). The corresponding cohort implicit (unknown) sample weights are $\left\{w_i^{(c)} = 1/\pi_i^{(c)}, i \in s_c\right\}$. In the cohort, we observe disease incidence and mortality with follow-up times and covariates: $\{D_i, X_i, \widetilde{D}_i, \widetilde{X}_i, \mathbf{z}_i, i \in s_c\}$. In addition, we assume there is a reference survey sample $s_s$ of $n_s$ individuals that are randomly sampled from the $FP$, with positive selection probabilities, i.e., $\pi_i^{(s)} = E_s\left(\delta_i^{(s)}\right) > 0$, for all $i \in FP$, where $\delta_i^{(s)}$ is the sample inclusion indicator (= 1 if $i \in s_s$; 0 otherwise), and $E_s$ is the expectation over all possible random samples according to the survey sample design. The corresponding sample weights are denoted by $w_i^{(s)} = 1/\pi_i^{(s)}$. In survey sample, we know variables $\{w_i^{(s)}, \widetilde{D}_i, \widetilde{X}_i, \mathbf{z}_i, i \in s_s\}$. In "design-based" finite population inference, the $FP$ qunataties $\boldsymbol{\beta}^{FP}$, $\Lambda_0^{FP}(t)$, and $r^{FP}$ can be treated as having small-order variability. The randomness mainly comes from the unknown cohort participation and survey sampling indicators $\left\{\delta_i^{(c)}, \delta_i^{(s)}, i \in FP\right\}$ (Elliott and Valliant, 2017).

## 3   Calibrated Pseudoweighting Methods with New Auxiliary Variables

We propose two calibrated pseudoweighting methods to estimate $\boldsymbol{\beta}^{FP}$, $\Lambda_0^{FP}(t)$, and $r^{FP}$ from the cohort $s_c$, by treating the combined (pseudo)weighted cohort and the reference survey sample $s_c \cup s_s$ as being representative of the $FP$, and treating $s_c$ as a subsample of $s_c \cup s_s$ in five steps. First, we create IPW pseudoweights for individuals in $s_c$ (Wang et al., 2021). Second, we combine the $s_c$ and $s_s$ with scaled weights so that the sum of the scaled weights in $s_c \cup s_s$ equals $M$ (i.e., number of individuals in the $FP$). Third, we create innovative auxiliary variables of influence functions for individuals in $s_c \cup s_s$. Fourth, we calibrate IPW pseudoweighted $s_c$ to the (pseudo) weighted $s_c \cup s_s$, using the auxiliary variables generated in the third step. The pure risks are finally estimated from $s_c$ using the calibrated pseudoweights. Please refer to Figure 1 for the flow chart of the proposed methods and we elaborate each step below in Sections 3.1 – 3.3.

Figure 1 Steps of calibrated pseudoweighting methods for risk estimation

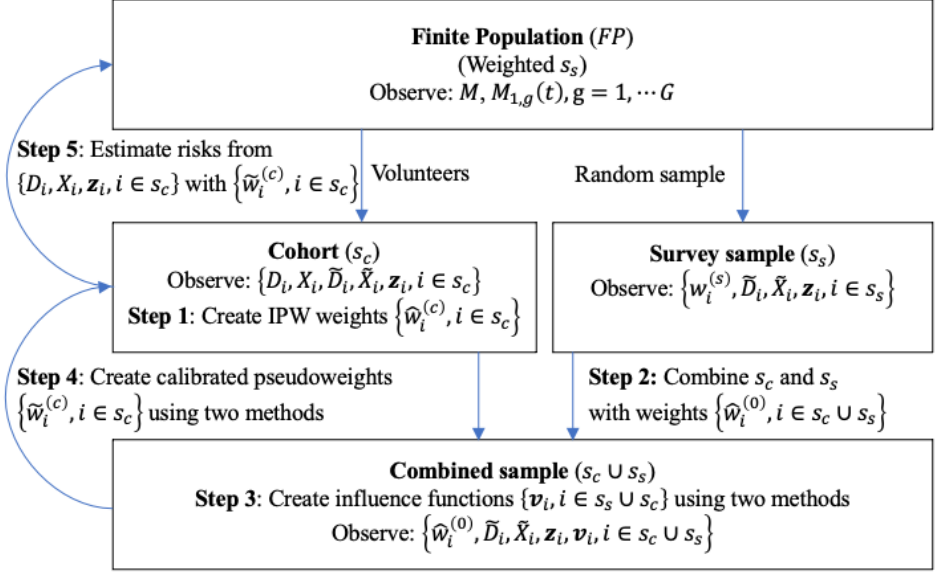

## 3.1 Inverse-Propensity Weighting (IPW) Method (Step 1 in Figure 1)

The IPW method estimates the implicit cohort sample weights $\{w_i^{(c)}, i \in s_c\}$. There are multiple existing IPW methods (Chen et al., 2020; Elliott and Valliant, 2017; Wang et al., 2021). Many of these methods have similar performance especially when the cohort sample fraction (i.e., $n_c/N$) is small. In this paper, we choose the IPW in Wang et al., 2021 as an example because it is easy to implement. Other IPW methods yield similar results and therefore are not shown in this paper. We refer the reader to Wang et al. (2021) for proof of consistency, and comparison of several commonly used IPW methods, which is not the focus of this paper. The IPW weights for cohort individuals are denoted by $\{\widehat{w}_i^{(c)}, i \in s_c\}$. Analogous to Wang et al. (2022), it can be proved that the IPW estimators of the HR's, baseline hazards, and the pure risks are consistent to the $FP$ quantities under the correctly specified propensity model and some other standard assumptions. In practice, the true propensity model is unknown. The IPW estimators can be biased under a misspecified propensity model. To obtain more bias reduction for risk estimation, the disease mortality indicator $\widetilde{D}_i$, and interactions between $z_i$ and $\widetilde{D}_i$ can also be included as predictors of $\Pr(R_i = 1)$ in the propensity model (see Section 4 for details).

## 3.2 Calibrate the IPW Cohort to the Combined Cohort and Survey Sample Using Influence Functions (Steps 2-4 in Figure 1)

The IPW cohort risk estimators can be sensitive to misspecified propensity models and inefficient due to highly variable weights especially for minorities. Calibration methods (Deville and Sarndal, 1992) have been widely used in survey research to reduce under coverage bias in Horvitz-Thompson inverse probability weighting estimators of totals and means of outcomes by borrowing strength from summary statistics of auxiliary variables in the $FP$ that are correlated with the outcome. Because summary statistics of the auxiliary variable in the $FP$ can be treated as having small-order variability, the calibrated estimation is also more efficient than the uncalibrated sample weighted estimation, when the auxiliary variable is highly correlated with the outcome.

Survey calibration methods can also be applied to cohort psuedoweights and calibrate the pseudoweighted cohort to the sample weighted reference survey sample on common covariates $z$ in the two samples to further reduce bias due to inaccurate pseudoweights (Lee et al., 2022). However, the existing calibration method can have limited bias reduction and harm the efficiency of the risk estimation for two reasons. First, covariates $z_i$ can be less correlated with risk estimates even though they can be associated with the outcome. Second, as the survey sample size is usually small, the calibration total from the weighted survey sample, $\sum_{i \in s_s} w_i^{(s)} z_i$, although consistent to the $FP$ total, can be highly variable. To overcome these two problems, we propose to calibrate the IPW pseudoweighted cohort to the combined cohort and survey sample (instead of survey sample only in the existing literature) on innovative auxiliary variables of influence functions (instead of covariates in standard survey calibration methods) which requires to leverage the individual cancer mortality data from the surveys.

When combining the weighted survey sample and the IPW pseudoweighted cohort (Step 2 in Figure 1), we use the scalar $a_d$, for $d = c, s$, calculated by (Korn and Graubard, 1999)

$$a_d = \frac{\sum_{i \in s_c \cup s_s} w_i^{(d)}}{2 \sum_{i \in s_d} w_i^{(d)}} \left\{ 1 - \frac{(CV_d^2 + 1)/n_d}{(CV_s^2 + 1)/n_s + (CV_c^2 + 1)/n_c} \right\}, \text{ for } d = c, s,$$

where we substitute the IPW pseudoweight $\widehat{w}_i^{(c)}$ for $w_i^{(c)}$. We denote the weights for the combined sample $s_c \cup s_s$ as $\{\widehat{w}_i^{(0)}, i \in s_c \cup s_s\}$ where $\widehat{w}_i^{(0)} = a_c \cdot \widehat{w}_i^{(c)}$ for $i \in s_c$, and $\widehat{w}_i^{(0)} = a_s \cdot w_i^{(s)}$ for $i \in s_s$. In the combined sample, we have variables $\{\widehat{w}_i^{(0)}, \mathbf{z}_i, \widetilde{D}_i, \widetilde{X}_i, i \in s_c \cup s_s\}$. Again, under the correctly specified propensity model, $s_c \cup s_s$ with the weights $\widehat{w}_i^{(0)}$ should represent the $FP$.

In Steps 3 and 4, we calibrate $s_c$ with the IPW pseudoweights $\{\widehat{w}_i^{(c)}, i \in s_c\}$ to $s_c \cup s_s$ with weights $\{\widehat{w}_i^{(0)}, i \in s_c \cup s_s\}$ using innovative calibration auxiliary variables of influence functions. The cohort calibrated pseudoweights, denoted by $\{\widetilde{w}_i^{(c)}, i \in s_c\}$, is obtained by minimizing the averaged distance between $\{\widetilde{w}_i^{(c)}, i \in s_c\}$ and $\{\widehat{w}_i^{(c)}, i \in s_c\}$ with the constraint $\sum_{i \in s_c} \widetilde{w}_i^{(c)} \mathbf{v}_i = \sum_{i \in s_c \cup s_s} \widehat{w}_i^{(0)} \mathbf{v}_i$, where $\{\mathbf{v}_i, i \in s_c \cup s_s\}$ is a set of proposed auxiliary variables that are constructed from the common variables in the combined sample $\{\widehat{w}_i^{(0)}, \mathbf{z}_i, \widetilde{D}_i, \widetilde{X}_i, i \in s_c \cup s_s\}$ (see details later in section 3.2). This is equivalent to solving the following equation:

$$Q(\boldsymbol{\eta}) = \sum_{i \in s_c} \widehat{w}_i^{(c)} F(\mathbf{v}_i^T \boldsymbol{\eta}) \mathbf{v}_i - \sum_{i \in s_c \cup s_s} \widehat{w}_i^{(0)} \mathbf{v}_i = \mathbf{0}, \tag{3.1}$$

where $F(\cdot)$ is the calibration factor determined by the selected distance function (Deville and Sarndal, 1992), and $\boldsymbol{\eta}$ is the nuisance parameter. When the chi-squared distance (Deville and Sarndal, 1992) is used, we have $F(v_i^T \eta) = v_i^T \eta + 1$, and $\hat{\eta}$ i.e., the solution of equation (3.1), has the following explicit form:

$$\hat{\eta} = \left\{ \sum_{i \in s_c} \hat{w}_i^{(c)} v_i^T v_i \right\}^{-1} \left\{ \sum_{i \in s_c \cup s_s} \hat{w}_i^{(0)} v_i - \sum_{i \in s_c} \hat{w}_i^{(c)} v_i \right\}, \quad (3.2)$$

The resulting calibration pseudoweight is $\widetilde{w}_i^{(c)} = F(v_i^T \hat{\eta}) \cdot \hat{w}_i^{(c)}$ for $i \in s_c$.

The auxiliary variables $\{v_i, i \in s_c \cup s_s\}$ should be highly correlated with sample estimators of $\beta^{FP}$ and $\Lambda_0^{FP}(t)$. If knowing the true sample weights, the outcome of disease incidence, and covariates in the combined sample, i.e., $\{w_i^{(0)}, z_i, D_i, X_i, i \in s_c \cup s_s\}$, we can estimate $\beta^{FP}$ from $s_s \cup s_s$, and the resulting estimator, denoted by $\hat{\beta}$, satisfies

$$\hat{\beta} - \beta = \sum_{i \in s_c \cup s_s} w_i^{(0)} \Delta_i(\hat{\beta}) + o\{(n_c + n_s)^{-1/2}\}, \quad (3.3)$$

where $\beta$ is the true value of $\beta^{FP}$ in the $FP$ estimating equation (2.2), and $\Delta_i(\hat{\beta})$ is the influence functions of $\hat{\beta}$, which is a function of the unknown true $\beta$. The sample plug-in influence functions of $\hat{\beta}$ are obtained by $\Delta(\hat{\beta}) = \{\Delta_i(\hat{\beta}) = \partial \hat{\beta}/\partial w_i^{(0)}, i \in s_c \cup s_s\}$. Please see Supplementary Material A.1 for justification of Equation (3.3) and derivation of $\Delta(\hat{\beta})$. Equation (3.3) implies that $\Delta(\hat{\beta})$ can be highly correlated with $\hat{\beta}$. Therefore, the auxiliary variables $\{v_i, i \in s_c \cup s_s\}$ for estimating $\beta^{FP}$ should include $\Delta(\hat{\beta})$ or variables that are strongly associated with $\Delta(\hat{\beta})$.

The cumulative baseline hazard $\Lambda_0^{FP}(t)$ can be consistently estimated from $s_c \cup s_s$ by

$$\hat{\Lambda}_0(t) = \int_0^t \frac{\sum_{i \in s_c \cup s_s} w_i^{(0)} dN_i(\tau)}{\sum_{i \in s_c \cup s_s} w_i^{(0)} Y_i(\tau) \exp(\hat{\beta}^T z_i)}, \quad (3.4)$$

if $\{w_i^{(0)}, z_i, D_i, X_i, i \in s_c \cup s_s\}$ is available. The numerator and the denominator of $\hat{\Lambda}_0(t)$ are respectively strongly associated with $\{D_i, i \in s_c \cup s_s\}$ and $\{X_i \exp(\hat{\beta}^T z_i), i \in s_c \cup s_s\}$, which can be included in the auxiliary variables for estimating $\Lambda_0^{FP}(t)$.

Importantly, because $(D_i, X_i)$ is **not** available in $s_s$, $\boldsymbol{\beta}^{FP}$ cannot be estimated from $s_c \cup s_s$. Therefore, we propose two methods that use surrogates of $\{\Delta_i(\widehat{\boldsymbol{\beta}}), D_i, X_i \exp(\widehat{\boldsymbol{\beta}}^T \mathbf{z}_i), i \in s_c \cup s_s\}$ as calibration auxiliary variables in Sections 3.2.1 and 3.2.2.

*3.2.1 Method 1 – Calibrated Inverse-Propensity Weighting Method (CIPW) Using Influence Functions of Log-HR's for Mortality (Steps 3-4 in Figure 1)*

Because disease-specific mortality can be strongly associated with disease incidence, the log-HR of disease-specific mortality in the $FP$, denoted by $\widetilde{\boldsymbol{\beta}}^{FP}$, will be associated with $\boldsymbol{\beta}^{FP}$. Given that $(\widetilde{D}_i, \widetilde{X}_i)$ is available in both $s_c$ and $s_s$, we propose to include influence functions of $\widehat{\widetilde{\boldsymbol{\beta}}}$ estimated from (pseudo)weighted $s_c \cup s_s$ as the auxiliary variables in $\boldsymbol{v}_i$ in formula (3.1). Specifically, we first fit a Cox regression model for disease-specific mortality, $(\widetilde{D}_i, \widetilde{X}_i)$, on predictors $\mathbf{z}_i$ in $s_c \cup s_s$ with pseudoweights $\{\widehat{w}_i^{(0)}, i \in s_c \cup s_s\}$. Then $\widetilde{\boldsymbol{\beta}}^{FP}$ can be estimated by solving the pseudo-estimating equation:

$$\widetilde{U}(\widetilde{\boldsymbol{\beta}}) = M^{-1} \sum_{i \in s_c \cup s_s} \int_0^\infty \widehat{w}_i^{(0)} \left\{ \mathbf{z}_i - \hat{\widetilde{S}}^{(0)}(\tau, \widetilde{\boldsymbol{\beta}})^{-1} \hat{\widetilde{S}}^{(1)}(\tau, \widetilde{\boldsymbol{\beta}}) \right\} d\widetilde{N}_i(\tau) = \mathbf{0}, \qquad (3.5)$$

where $\hat{\widetilde{S}}^{(u)}(\tau, \widetilde{\boldsymbol{\beta}}) = M^{-1} \sum_{j \in s_c \cup s_s} \widehat{w}_i^{(0)} \widetilde{Y}_j(\tau) \exp(\widetilde{\boldsymbol{\beta}}^T \mathbf{z}_j) \cdot \mathbf{z}_j^{\otimes u}$. Denote the solution of (3.5) as $\widehat{\widetilde{\boldsymbol{\beta}}}$. The influence function of $\widehat{\widetilde{\boldsymbol{\beta}}}$ is

$$\widehat{\Delta}_i\left(\widehat{\widetilde{\boldsymbol{\beta}}}\right) = \partial \widehat{\widetilde{\boldsymbol{\beta}}} / \partial \widehat{w}_i^{(0)} = M^{-1} \widetilde{U}_{\widetilde{\boldsymbol{\beta}}}^{-1} \int_0^\infty \left\{ \mathbf{z}_i - \hat{\widetilde{S}}^{(0)}\left(\tau, \widehat{\widetilde{\boldsymbol{\beta}}}\right)^{-1} \hat{\widetilde{S}}^{(1)}\left(\tau, \widehat{\widetilde{\boldsymbol{\beta}}}\right) \right\} d\widetilde{M}_i(\tau) \qquad (3.6)$$

where $\widetilde{U}_{\widetilde{\boldsymbol{\beta}}} = \partial \widetilde{U}(\widetilde{\boldsymbol{\beta}})/\partial \widetilde{\boldsymbol{\beta}}$, and $d\widetilde{M}_i(\tau) = d\widetilde{N}_i(\tau) - \hat{\widetilde{S}}^{(0)}(\tau, \widetilde{\boldsymbol{\beta}})^{-1} Y_i(\tau) \exp(\widetilde{\boldsymbol{\beta}}^T \mathbf{z}_i) \sum_{k \in s_c \cup s_s} \widehat{w}_k^{(0)} \cdot d\widetilde{N}_k(\tau)$ (details in Supplementary materials A.2).

We use the auxiliary variables $\boldsymbol{v}_i = \left\{1, \widetilde{D}_i, \widehat{\Delta}_i\left(\widehat{\widetilde{\boldsymbol{\beta}}}\right)\right\}$ in equations (3.1) and (3.2) to obtain the calibration adjustment factors for the IPW pseudoweights to estimate $\boldsymbol{\beta}^{FP}$, denoted by $\{F_{\boldsymbol{\beta},i}^{CIPW}, i \in s_c\}$. Similarly, we use auxiliary variables $\boldsymbol{v}_i = \left\{1, \widetilde{D}_i, \widetilde{X}_i \exp\left(\widehat{\widetilde{\boldsymbol{\beta}}}^T \mathbf{z}_i\right)\right\}$ for calibration of the IPW pseudoweights to estimate $\Lambda_0^{FP}(t)$. Denote the corresponding calibration adjustment factor as $\{F_{\lambda,i}, i \in s_c\}$.

### 3.2.2 Method 2 – Calibrated Inverse-Propensity Weighting Method Using Influence Function of Log-HR's for Incidence with Imputation (CIPW-I) (Steps 3-4 in Figure 1)

For less fatal diseases, such as breast cancer, using auxiliary variables $\boldsymbol{v}_i = \left\{1, \widetilde{D}_i, \widehat{\Delta}_i\left(\widehat{\widetilde{\boldsymbol{\beta}}}\right), \widetilde{X}_i \exp\left(\widehat{\widetilde{\boldsymbol{\beta}}}^T \boldsymbol{z}_i\right), i \in s_c \cup s_s\right\}$ generated from the disease-specific mortality model may result in limited bias reduction or efficiency gain for two reasons. First, the association between disease-specific mortality and disease incidence may not be strong for less fatal diseases. Second, many cases living with the cancer in the cohort are treated as censored in the mortality estimating equation (3.5) and cannot be used for calibration. To try to mitigate this issue, we consider using auxiliary variables generated from the Cox regression model for disease incidence fitted to the combined sample $s_c \cup s_s$, with the imputed time-to-incidence for mortality cases in $s_s$. The imputation steps are as follows:

(1) Calculate the gap time between disease incidence and mortality, $\Delta T_i = \widetilde{T}_i - T_i$, for mortality cases in the cohort, i.e., $\{i, i \in s_c, \widetilde{D}_i = 1\}$, and fit an imputation model for $\Delta T_i$ using covariates $\boldsymbol{z}$ to the disease-specific mortality cases in the cohort with the IPW pseudoweights.

(2) For mortality cases in the survey sample, i.e. $\{i, i \in s_s, \widetilde{D}_i = 1\}$, impute the gap time, $\Delta T_i^* > 0$ from the imputation model developed in above step (1), and impute the incidence time by $T_i^* = \max(0, \widetilde{T}_i - \Delta T_i^*)$.

(3) Create the (imputed) incidence indicator, and observed incidence time, $(D_i^*, X_i^*)$, in the combined sample $s_c \cup s_s$ as follows:

$$(D_i^*, X_i^*) = \begin{cases} (D_i, X_i) \text{ for } i \in s_c \\ (\widetilde{D}_i, X_i^*) \text{ for } i \in s_s \end{cases},$$

where $X_i^* = T_i^*$ for $\widetilde{D}_i = 1$, and $X_i^* = X_i$ otherwise.

We then fit a Cox regression model for $(D_i^*, X_i^*)$ on covariates $\mathbf{z}_i$ in $s_c \cup s_s$ with weights $\{\widehat{w}_i^{(0)}, i \in s_c \cup s_s\}$. The log-HR $\boldsymbol{\beta}^{FP}$ can be estimated by solving the pseudo-estimating equation:

$$\widehat{U}^*(\boldsymbol{\beta}) = \sum_{i \in s_c \cup s_s} \int_0^\infty \widehat{w}_i^{(0)} \{\mathbf{z}_i - \widehat{S}^{*(0)}(\tau, \boldsymbol{\beta})^{-1} \widehat{S}^{*(1)}(\tau, \boldsymbol{\beta})\} dN_i^*(\tau) = \mathbf{0}, \qquad (3.7)$$

Where $\widehat{S}^{*(u)}(\tau, \boldsymbol{\beta}) = M^{-1} \sum_{j \in s_c \cup s_s} \widehat{w}_i^{(0)} Y_j^*(\tau) \exp(\boldsymbol{\beta}^T \mathbf{z}_j) \cdot \mathbf{z}_j^{\otimes u}$, with $Y_i^*(t) = 1$ if $X_i^* \geq t$; and $Y_i^*(t) = 0$ otherwise. Denote the solution of (3.7) as $\widehat{\boldsymbol{\beta}}^*$. Accordingly, the calibration adjustment factors for estimating $\boldsymbol{\beta}^{FP}$, denoted by $F_{\boldsymbol{\beta},i}^{CIPW-I}$, and for $\lambda_0(t)$, denoted by $F_{\lambda,i}^{CIPW-I}$, are calculated from equations (3.1) and (3.2) with the auxiliary variables being $\mathbf{v}_i = \{1, D_i^*, \Delta_i(\widehat{\boldsymbol{\beta}}^*)\}$ and $\mathbf{v}_i = \{1, D_i^*, X_i^* \exp(\widehat{\boldsymbol{\beta}}^{*T} \mathbf{z}_i)\}$ respectively, with $\Delta_i(\widehat{\boldsymbol{\beta}}^*) = \partial \widehat{\boldsymbol{\beta}}^* / \partial \widehat{w}_i^{(0)}$.

Finally, if the composite incidence rates by age/sex/race groups in the $FP$ are also available, both the CIPW and the CIPW-I calibrated pseudoweights can be poststratified to the $FP$ so that the group-specific incidence rates in the pseudoweighted cohort are approximately the same with that in the $FP$. The poststratification step has been shown to improve accuracy and efficiency in estimating $\boldsymbol{\beta}^{FP}$ for age, sex, and race/ethnicity, $\Lambda_0^{FP}$, and $r^{FP}$ (Wang et al., 2022).

### 3.3 CIPW and CIPW-I Estimators of $\boldsymbol{\beta}^{FP}$, $\Lambda_0^{FP}$, and $r^{FP}$ (Step 5 in Figure 1)

The CIPW estimator of $\boldsymbol{\beta}^{FP}$, denoted by $\widehat{\boldsymbol{\beta}}^{CIPW}$, is estimated by solving the calibrated pseudo estimating equation for $\boldsymbol{\beta}$

$$\widehat{U}(\boldsymbol{\beta}) = \sum_{i \in s_c} \int_0^\infty F_{\boldsymbol{\beta},i}^{CIPW} w_i^{(0)} \{\mathbf{z}_i - \widehat{S}^{(0)}(\tau, \boldsymbol{\beta})^{-1} \widehat{S}^{(1)}(\tau, \boldsymbol{\beta})\} dN_i(\tau) = \mathbf{0}, \qquad (3.8)$$

where $\widehat{S}^{(u)}(\tau, \boldsymbol{\beta}) = M^{-1} \sum_{j \in s_c} F_{\boldsymbol{\beta},i}^{CIPW} \widehat{w}_i^{(0)} Y_j(\tau) \exp(\boldsymbol{\beta}^T \mathbf{z}_j) \cdot \mathbf{z}_j^{\otimes u}$, $u = 0, 1$. The CIPW estimator of $\Lambda_0^{FP}(t)$ is obtained by

$$\widehat{\Lambda}_0^{CIPW}(t) = \int_0^t \{1 - \widehat{AR}(\tau, \widehat{\boldsymbol{\beta}}^{CIPW})\} \cdot \lambda_{FP}(\tau) \, d\tau, \tag{3.9}$$

where $\widehat{AR}(\tau, \widehat{\boldsymbol{\beta}}^{CIPW}) = 1 - \hat{S}^{(0)}(\tau, \widehat{\boldsymbol{\beta}}^{CIPW})^{-1} \hat{S}^{(0)}(\tau)$ is the CIPW estimator of the "population attributable risk" (Gail, 1989), with $\hat{S}^{(0)}(\tau) = M^{-1} \sum_{i \in S_c} F_{\lambda,i}^{CIPW} \widehat{w}_i^{(c)} Y_i(\tau)$ being the CIPW estimator of proportion of risk set in the $FP$, and $\lambda_{FP}(\tau) = \{\sum_{i \in FP} Y_i(\tau)\}^{-1} \{\sum_{i \in FP} dN_i(\tau)\}$ is the composite disease incidence rate at each incidence time $\tau$ that can be calculated from the population registry. Compared with the Breslow cumulative baseline hazard estimator, this population attributable risk (PAR) estimator can have smaller variance due to borrowing composite event rates from the $FP$ (Wang et al., 2022). Based on the estimating equation theory, it can be shown that $\widehat{\boldsymbol{\beta}}^{CIPW}$ and $\widehat{\Lambda}_0^{CIPW}(t)$ are consistent estimators of $\boldsymbol{\beta}^{FP}$ and $\Lambda_0^{FP}(t)$ respectively under the correct propensity model and other standard conditions (Supplementary Materials B). Furthermore, $\widehat{\boldsymbol{\beta}}^{CIPW}$ and $\widehat{\Lambda}_0^{CIPW}(t)$ can be more efficient and less sensitive to misspecified propensity model than the IPW estimators especially when the disease of interest is highly fatal (e.g., lung cancer). This is because the calibration step borrows strength of the extra disease mortality cases in the survey sample, which is strongly associated with the disease incidence.

Finally, the CIPW pure risk estimator is:

$$\hat{r}^{CIPW}(t, \boldsymbol{z}) = 1 - \exp\{-\widehat{\Lambda}_0^{CIPW}(t) \cdot \exp(\boldsymbol{z}^T \widehat{\boldsymbol{\beta}}^{CIPW})\}, \tag{3.10}$$

The CIPW-I estimators of $\boldsymbol{\beta}^{FP}$, $\Lambda_0^{FP}(t)$, and $r^{FP}$, respectively denoted by $\widehat{\boldsymbol{\beta}}^{CIPW\text{-}I}$, $\widehat{\Lambda}_0^{CIPW\text{-}I}(t)$, and $\hat{r}^{CIPW}(t, \boldsymbol{z})$, are obtained by changing $F_{\boldsymbol{\beta},i}^{CIPW}$ in equation (3.8) to $F_{\boldsymbol{\beta},i}^{CIPW-I}$, changing $F_{\lambda,i}^{CIPW}$ and $\widehat{\boldsymbol{\beta}}^{CIPW}$ in formula (3.9) to $F_{\lambda,i}^{CIPW-I}$ and $\widehat{\boldsymbol{\beta}}^{CIPW\text{-}I}$, and changing $\widehat{\boldsymbol{\beta}}^{CIPW}$ and $\widehat{\Lambda}_0^{CIPW}(t)$ in formula (3.10) to $\widehat{\boldsymbol{\beta}}^{CIPW-I}$ and $\widehat{\Lambda}_0^{CIPW-I}(t)$ respectively.

## 3.4 Variance Estimation

As it is very complicated to derive the analytical plug-in variance estimators shown in Supplementary Materials B, we apply the jackknife replication method for variance estimation to account for all sources of variability due to cohort participation (including propensity estimation and calibration), survey sample selection, calibration, and imputation. The detailed steps of the jackknife variance estimation are described in Supplementary materials C. Briefly, we leave out one primary sampling unit (PSU) in the survey sample or one random group of individuals in the cohort and assign replicate weights for the remaining sample (Valliant et al., 2013; Wang et al., 2020). Then we follow steps 1-5 in Figure 1 to create calibrated psuedoweights for the remaining cohort by weighting the observations in the remaining $s_c \cup s_s$ with the jackknife replicate weights. The pure risk is estimated for each replicate. Coverage probabilities by using our jackknife variance are close to nominal level (see Section 4.2).

## 4 SIMULATION STUDIES

### 4.1 Finite population, registry (vital statistics), cohort and survey sample generation

We generated a finite population $FP$ of size $M = 300,000$ with three covariates $z_1 \sim N(0,4)$, $z_2 \sim N(0,2)$, and $z_3 \sim N(0,2)$. We generated time to event using a $Weibull(\theta, \alpha = 1)$ with $\theta = \exp(\beta_0 + \boldsymbol{\beta}^T \boldsymbol{z})$, where $\boldsymbol{\beta} = (\beta_1, \beta_2, \beta_3)^T$, and $\boldsymbol{z} = (z_1, z_2, z_3)^T$. The hazard function $\lambda(t; \boldsymbol{z}) = \lambda_0(t) \exp(\boldsymbol{\beta}^T \boldsymbol{z})$ is time-invariant, with baseline hazard $\lambda_0(t) = \exp(\beta_0)$ and pure risk is $r(t; \boldsymbol{z}) = 1 - \exp\left\{-\int_0^t \exp(\beta_0) \exp(\boldsymbol{\beta}^T \boldsymbol{z})\right\}$. We set $\beta_0 = \log\{-(\log 0.85)/15\}$, and $\boldsymbol{\beta} = (0.2, 0.2, 0.3)$. We set administrative censoring at 15 years after start of the study and $C_0 = 15 - T_0$ was the time from entry to administrative censoring, where $T_0 \sim U(0,1)$ is the random starting time. We considered censoring from death due to other causes, $C \sim Weibull\{\theta = -(\log 0.9)/15, \alpha = 1\}$. The observed time variable is $X = \min(T, C_0, C)$, with $T \sim Weibull\{\theta, \alpha = 1\}$ as the true disease incidence time, and $D$ indicating the disease incidence of interest, $D = I\{T \leq \min(C_0, C)\}$. The 15-year incidence rate in the $FP$ was $M_1/M = 19\%$. The gap time between disease incidence and disease-specific mortality is a truncated normal distribution $T_\Delta = \max(0, \beta_1^* z_1 + \beta_2^* z_2 + \beta_{1,2}^* z_1 z_2 + \epsilon)$, with $\epsilon \sim N(\mu, \sigma)$. Values of $\boldsymbol{\beta}^* = (\beta_0^*, \beta_1^*, \beta_2^*, \beta_{1,2}^*)^T$, $\mu$, and $\sigma$ were varied in 3 scenarios for disease lethality and the association between the gap time and disease predictors:

(1) Fatal disease, mild association: $\beta_0^* = 2, \beta_1^* = \beta_2^* = \beta_{1,2}^* = 0.01, \mu = 2, \sigma = 0.01$

(2) Less fatal disease, moderate association: $\beta_1^* = \beta_2^* = \beta_{1,2}^* = 0.1\ \mu = 10, \sigma = 0.2$

(3) Less fatal disease, no association: $\beta_1^* = \beta_2^* = \beta_{1,2}^* = 0\ \mu = 10, \sigma = 0.2$

The disease-specific mortality rate was 17.3%, 5.8%, and 7.0% in the 3 scenarios respectively. The $FP$ parameters $\boldsymbol{\beta}^{FP}$, $\Lambda_0^{FP}(t)$, and $r^{FP}$ are every close to the true values $\boldsymbol{\beta}$, $\Lambda_0(t)$, and $r$ respectively.

A cohort of $n_c = 3{,}000$ was randomly selected using Probability Proportional to Size (PPS) sampling with the measure of size defined by $\exp(\gamma_1 z_1 + \gamma_2 z_2 + \gamma_d D + \gamma_{2,d} z_2 \cdot D)$ where $\gamma_1 = -0.15, \gamma_2 = 0.1$. The values of $\gamma_d$ and $\gamma_{2,d}$ were $\gamma_d = \gamma_{2,d} = 0$ and $\gamma_d = -0.75, \gamma_{2,d} = -0.2$ for noninformative and informative cohort selection respectively. Although we generated random samples of cohorts, cohort recruitment is usually volunteer based, with unobserved cohort participation rates in practice. Hence, in the analysis, we masked the PPS design and treated the cohort as a simple random sample for naïve risk estimation, as is typical in cohort analyses. A survey of size $n_s = 6{,}000$ was randomly selected using PPS sampling with selection probability $\exp(0.7 z_1 + 0.7 z_2)$.

## 4.2 Simulation results

Results for the 3 disease lethality scenarios under noninformative or informative cohort participation are in Table 2 and Table 3. Notice that we fitted a linear regression model including all $\boldsymbol{z}$ as predictors to estimate gap time between disease incidence and disease-specific mortality time in the imputation step for the CIPW-I method.

Table 2 Estimates of Log-odds Ratios in Three Scenarios of Simulations

| Method | Relative Bias (%) | | | Variance ($\times 10^{-4}$) | | | MSE ($\times 10^{-4}$) | | | 95% CP (JK) | | |
|---|---|---|---|---|---|---|---|---|---|---|---|---|
| | $\beta_1$ | $\beta_2$ | $\beta_3$ | $\beta_1$ | $\beta_2$ | $\beta_3$ | $\beta_1$ | $\beta_2$ | $\beta_3$ | $\beta_1$ | $\beta_2$ | $\beta_3$ |
| **Noninformative cohort participation** $\pi_i^{(c)} \propto \exp(-0.15z_1 + 0.1z_2)$ | | | | | | | | | | | | |
| Naive | 0.3 | 0.0 | 0.2 | 1.56 | 6.15 | 6.12 | 1.57 | 6.15 | 6.13 | 0.94 | 0.94 | 0.95 |
| **Fitted propensity model** $\text{logit}(p) \sim z_1 + z_2$ | | | | | | | | | | | | |
| IPW | 0.5 | -0.3 | 0.6 | 3.10 | 9.47 | 8.66 | 3.11 | 9.47 | 8.70 | 0.94 | 0.94 | 0.95 |
| Scenario 1 | | | | | | | | | | | | |
| CIPW | 0.2 | -0.1 | 0.2 | 0.89 | 2.69 | 2.63 | 0.89 | 2.69 | 2.63 | 0.94 | 0.95 | 0.95 |
| CIPW-I | -0.2 | -1.0 | -0.6 | 0.65 | 1.93 | 1.98 | 0.65 | 1.98 | 2.00 | 0.94 | 0.95 | 0.94 |
| Scenario 2 | | | | | | | | | | | | |
| CIPW | 0.4 | -0.2 | 0.3 | 2.62 | 6.93 | 6.52 | 2.62 | 6.93 | 6.53 | 0.94 | 0.94 | 0.95 |
| CIPW-I | -13.7 | -25.1 | -0.2 | 1.51 | 4.26 | 4.20 | 8.87 | 29.46 | 4.21 | 0.42 | 0.32 | 0.95 |
| Scenario 3 | | | | | | | | | | | | |
| CIPW | 0.2 | -0.3 | 0.3 | 2.16 | 6.32 | 5.84 | 2.17 | 6.33 | 5.84 | 0.94 | 0.94 | 0.95 |
| CIPW-I | 0.6 | -0.5 | 0.3 | 1.31 | 3.85 | 3.86 | 1.33 | 3.86 | 3.87 | 0.94 | 0.95 | 0.94 |
| **Informative cohort participation** $\pi_i^{(c)} \propto \exp(-0.15z_1 + 0.1z_2 - 0.75D - 0.2z_2 \cdot D)$ | | | | | | | | | | | | |
| Naive | 8.5 | -83.5 | 8.1 | 3.73 | 13.6 | 14.8 | 6.55 | 292.9 | 20.7 | 0.87 | 0.01 | 0.90 |
| **Fitted propensity model** $\text{logit}(p) \sim z_1 + z_2 + \widetilde{D} + z_2 \cdot \widetilde{D}$ | | | | | | | | | | | | |
| Scenario 1 | | | | | | | | | | | | |
| IPW | 1.6 | -3.8 | 2.5 | 5.25 | 8.63 | 16.1 | 5.36 | 9.20 | 16.6 | 0.93 | 0.95 | 0.94 |
| CIPW | 0.5 | -5.6 | 0.5 | 1.02 | 3.14 | 2.93 | 1.03 | 4.42 | 2.96 | 0.95 | 0.89 | 0.95 |
| CIPW-I | 0.2 | -1.8 | -0.1 | 0.79 | 2.59 | 2.34 | 0.79 | 2.72 | 2.34 | 0.94 | 0.94 | 0.96 |
| Scenario 2 | | | | | | | | | | | | |
| IPW | 1.6 | -47.8 | 11.2 | 6.96 | 13.2 | 21.3 | 7.06 | 104.8 | 32.5 | 0.93 | 0.30 | 0.88 |
| CIPW | -6.1 | -53.5 | 3.6 | 3.33 | 8.93 | 9.10 | 4.77 | 123.5 | 10.2 | 0.87 | 0.06 | 0.93 |
| CIPW-I | -19.2 | -52.4 | 3.6 | 2.21 | 6.34 | 6.24 | 16.7 | 116.2 | 7.40 | 0.29 | 0.02 | 0.92 |
| Scenario 3 | | | | | | | | | | | | |
| IPW | 9.8 | -32.1 | 10.9 | 7.26 | 13.6 | 21.6 | 11.0 | 54.95 | 32.2 | 0.89 | 0.61 | 0.88 |
| CIPW | 3.2 | -38.0 | 3.8 | 2.69 | 8.23 | 7.98 | 3.10 | 66.16 | 9.30 | 0.93 | 0.28 | 0.93 |
| CIPW-I | 3.2 | -11.1 | 4.0 | 2.05 | 5.79 | 5.47 | 2.45 | 10.74 | 6.92 | 0.92 | 0.85 | 0.92 |

When cohort participation was noninformative (i.e., implicit cohort participation was only associated with $z$), the naïve, IPW and CIPW estimates of $\beta^{FP}$ were unbiased. The IPW estimates were the least efficient and the CIPW method, by calibrating the IPW weighted $s_c$ to $s_c \cup s_s$ on influence functions, substantially improved the efficiency of the IPW estimators when the disease was highly fatal (scenario 1). However, the efficiency improvement was limited for less fatal disease (scenarios 2 and 3). The CIPW-I estimators, by imputing the disease incidence time for disease-specific mortality cases in $s_s$, further improved the efficiency of the CIPW estimators especially when the disease is less fatal, and the gap time is not associated with the disease predictors (scenarios 3). Similar findings were found for the pure risk (Table 3)

Table 3 Estimates of Pure Risk at Three Time Point in Three Scenarios of Simulations

| Method | Relative Bias (%) | | | Variance ($\times 10^{-4}$) | | | MSE ($\times 10^{-4}$) | | | 95% CP (JK) | | |
|---|---|---|---|---|---|---|---|---|---|---|---|---|
| | t=1 | t=7 | t=15 | t=1 | t=7 | t=15 | t=1 | t=7 | t=15 | t=1 | t=7 | t=15 |
| **Noninformative cohort participation** $\pi_i^{(c)} \propto \exp(-0.15z_1 + 0.1z_2)$ | | | | | | | | | | | | |
| Naïve | 46.8 | 40.9 | 33.9 | 0.73 | 19.87 | 54.66 | 35.02 | 1169.2 | 3352.8 | 0.00 | 0.00 | 0.00 |
| **Fitted propensity model** $\text{logit}(p) \sim z_1 + z_2$ | | | | | | | | | | | | |
| IPW | 0.2 | 0.2 | -1.0 | 1.23 | 32.53 | 88.94 | 1.23 | 32.55 | 92.07 | 0.93 | 0.94 | 0.93 |
| Scenario 1 | | | | | | | | | | | | |
| CIPW | 0.3 | 0.2 | -1.1 | 0.43 | 9.99 | 29.19 | 0.43 | 10.01 | 32.81 | 0.95 | 0.95 | 0.94 |
| CIPW-I | 0.7 | 0.5 | -0.9 | 0.39 | 8.32 | 22.64 | 0.40 | 8.47 | 24.98 | 0.94 | 0.95 | 0.94 |
| Scenario 2 | | | | | | | | | | | | |
| CIPW | 0.0 | -0.1 | -1.3 | 0.96 | 26.85 | 74.02 | 0.96 | 26.86 | 78.64 | 0.94 | 0.95 | 0.94 |
| CIPW-I | 16.2 | 13.8 | 10.3 | 0.71 | 17.59 | 47.66 | 4.82 | 149.3 | 353.8 | 0.35 | 0.25 | 0.32 |
| Scenario 3 | | | | | | | | | | | | |
| CIPW | 0.1 | -0.1 | -1.3 | 0.75 | 22.66 | 62.58 | 0.75 | 22.66 | 67.24 | 0.94 | 0.94 | 0.94 |
| CIPW-I | 0.0 | -0.1 | -1.3 | 0.62 | 14.79 | 41.08 | 0.62 | 14.80 | 46.17 | 0.94 | 0.94 | 0.94 |
| **Informative cohort participation** $\pi_i^{(c)} \propto \exp(-0.15z_1 + 0.1z_2 - 0.75D - 0.2z_2 \cdot D)$ | | | | | | | | | | | | |
| Naïve | 82.6 | 67.9 | 51.8 | 1.07 | 28.29 | 77.22 | 107.90 | 31978 | 7799 | 0.00 | 0.00 | 0.00 |
| **Fitted propensity model** $\text{logit}(p) \sim z_1 + z_2 + \tilde{D} + z_2 \cdot \tilde{D}$ | | | | | | | | | | | | |
| Scenario 1 | | | | | | | | | | | | |
| IPW | 0.5 | 0.9 | -1.1 | 2.57 | 64.26 | 169.8 | 2.58 | 64.83 | 173.6 | 0.94 | 0.94 | 0.94 |
| CIPW | 0.1 | 0.4 | -1.4 | 0.71 | 12.20 | 34.41 | 0.71 | 12.32 | 40.20 | 0.94 | 0.94 | 0.94 |
| CIPW-I | -0.6 | -0.2 | -2.0 | 0.72 | 11.85 | 30.92 | 0.73 | 11.89 | 42.58 | 0.94 | 0.95 | 0.92 |
| Scenario 2 | | | | | | | | | | | | |
| IPW | 13.0 | 10.6 | 4.4 | 3.02 | 78.22 | 219.5 | 5.67 | 156.2 | 274.0 | 0.79 | 0.80 | 0.89 |
| CIPW | 13.8 | 12.6 | 7.4 | 1.28 | 37.90 | 108.6 | 4.26 | 147.1 | 268.0 | 0.64 | 0.60 | 0.76 |
| CIPW-I | 22.7 | 20.2 | 13.6 | 1.14 | 24.14 | 69.32 | 9.22 | 305.5 | 599.9 | 0.27 | 0.11 | 0.26 |
| Scenario 3 | | | | | | | | | | | | |
| IPW | 1.4 | 2.6 | -1.8 | 3.46 | 83.57 | 236.4 | 3.49 | 88.10 | 246.1 | 0.93 | 0.93 | 0.94 |
| CIPW   | 2.5  | 4.4  | 1.0  | 0.97 | 30.11 | 91.22 | 1.07 | 43.48 | 94.14 | 0.93 | 0.89 | 0.94 |
| CIPW-I | -4.1 | -0.8 | -3.5 | 1.17 | 20.84 | 66.30 | 1.43 | 21.26 | 101.4 | 0.93 | 0.94 | 0.89 |

However, when disease is less fatal and the gap time is associated with disease predictors $z$ (scenario 2), the CIPW-I estimators were biased even under the correct imputation model for the gap time. This is because only disease-specific mortality cases in $s_s$ can have imputed disease incidence times. Individuals who were alive with the disease in $s_s$ were censored at the end of follow-up time $C_0$. This censoring depends on both gap time between disease incidence and disease-specific mortality, $\Delta T$, and $C_0$. That is, individual $i \in s_s$ is censored if $T_i \geq C_0 - \Delta T_i$. In scenario 3, this censoring is the noninformative, (i.e., independent of survival) because $\Delta T$ is completely random. In scenario 2, however, the censoring is informative, because $\Delta T$ is associated with predictors $z$. As a result, CIPW-I estimators of $\boldsymbol{\beta}^{FP}$ are unbiased in scenario 3, but biased in scenario 2.

When the implicit cohort participation was informative, (i.e., associated with the disease incidence $D$ given the predictors $z$), the naïve cohort estimators of $\boldsymbol{\beta}^{FP}$ were substantially biased. Because disease incidence $D_i$ is not available in $s_s$, and was substituted by mortality status $\widetilde{D}_i$ in the propensity model. Due to the misspecified propensity model, all pseudoweighting methods mitigate, but do not eliminate, the bias. The bias reduction depended on the association between the disease incidence and disease-specific mortality, with the smallest bias for highly fatal disease (scenario 1), and highest bias for less fatal disease (scenario 2). Although bias cannot be eliminated, both the CIPW and the CIPW-I methods improved MSE in estimating $\boldsymbol{\beta}^{FP}$ and $r^{FP}$ compared to the IPW method, especially in scenarios 1 and 3 (Table 2, Table 3).

Based on the simulation results, both CIPW and CIPW-I are recommended when the disease of interest is fatal. When the disease of interest is less fatal, we recommend CIPW-I if the gap time between the disease incidence and disease-specific mortality is not associated with the disease predictors. Otherwise, CIPW is preferred.

## 5 PURE ESTIMATION FOR LUNG CANCER INCIDENCE

We estimate the pure risk of lung cancer incidence using PLCO, NHIS 1997-2000 as the reference survey sample, and obtained US cancer mortality and incidence rates from the CDC registry.

PLCO recruited $n_c = 149,905$ participants ages 55-74 years at 10 centers across the U.S from 1993-2001. Covariates are age, sex, race ethnicity, marital status, education, employment status, body mass index [BMI], smoking behaviors, and family history of cancer. We also created a region variable based on the location of the 10 PLCO centers. Participants were followed for cancer and mortality up to 2015 (median follow-up time 12.4 years).

The NHIS is a cross-sectional household interview survey of the non-institutionalized civilian US population. We choose the contemporaneous 1997-2000 NHIS respondents aged 55–74 years ($n_s = 30,204$ participants) as an appropriate reference for PLCO. The 1997 NHIS is a multistage stratified cluster sample with 339 strata each consisting of two sampled PSUs (NCHS, 2013). NHIS was also linked to the US National Death Index through 2006 for cause-specific mortality (NCHS, 2009). All links were presumed to be correct in this analysis. Importantly, no link has been made from NHIS to US cancer registries, and thus cancer incidence outcomes are unavailable in the NHIS.

Compared to the weighted NHIS, which is expected to represent the target population, the PLCO overrepresented males, Non-Hispanic Whites, younger people, higher education level, and those who are currently working. PLCO underrepresented current smokers and had a higher proportion of heavy smokers (Table S.1). Due to "healthy volunteer effects", lung cancer incidence rates in PLCO were consistently lower than that in the CDC registry, overall or by age group (Figure 2). In contrast, lung cancer mortality rates and distribution of age/sex/ race/ethnicity in the weighted NHIS, although variable, were close to the population, indicating that the NHIS sample was representative of lung cancer risk (Table S.1 and Figure S.1).

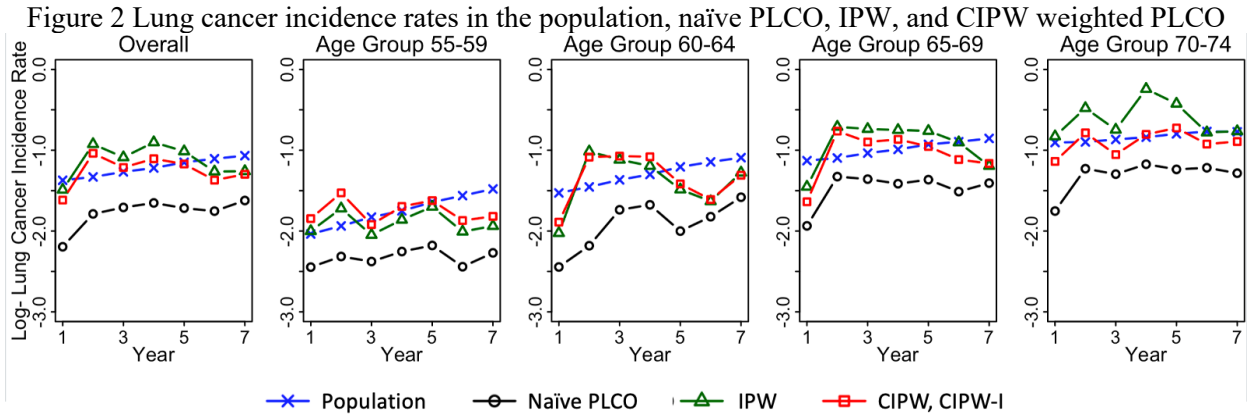

Figure 2 Lung cancer incidence rates in the population, naïve PLCO, IPW, and CIPW weighted PLCO

We used lung cancer incidence, and mortality rates and all-cause mortality rates 1999-2006 among people ages 55-74 in 1999 in the US (CDC 2018) to construct population composite incidence rates in the PAR method baseline cumulative hazard. Because lung cancer incidence is unavailable in NHIS, we used Centers for Disease Control and Prevention (CDC) lung cancer incidence rates by age, sex, and race/ethnicity groups as the "gold standard" to investigate the representativeness of the naïve PLCO, and to evaluate performance of the proposed methods.

The lung cancer incidence risk model included age, sex, race/ethnicity, education, BMI, and smoking: smoking status with quit years and cigarettes per day for former and current smokers respectively, smoking duration (in years) for ever smokers, and its interaction with current smoker. We then estimated the IPW weights for PLCO by fitting the propensity model to the combined PLCO and weighted NHIS 1997-2000. Following Li et al. (2022) we included common covariates in PLCO and NHIS that have different distributions in the naïve PLCO and weighted NHIS and are also associated with lung cancer incidence. We included all lung cancer incidence model covariates, as well as marital status, which was excluded from the mortality model as being inappropriate for clinicians to ask patients (Table S.2). Interactions of lung cancer mortality and age, sex, race, and smoking were also included as they are strongly associated with lung cancer incidence. For the CIPW-I method, we used linear regression model to estimate the gap time between lung cancer incidence and mortality, including all predictors in the risk model, at the imputation step. The CIPW and CIPW-I weights were poststratified to the population on lung cancer incidence rates by age/sex/race group.

IPW nearly equalized the covariate distribution in the PLCO to the weighted NHIS (Table S.1) and reduced bias in estimating lung cancer incidence rates (Figure 2). Moreover, as shown in Table 4, IPW yielded smaller estimates of HRs of lung cancer incidence for smoking behaviors than the naïve cohort, which are suspected to be overestimated, since PLCO had a higher proportion of current heavy smokers yet lower lung cancer incidence rates compared to NHIS or CDC. This conjecture is supported by the similar pattern observed in HRs for smoking behaviors in a model for lung cancer mortality (Table S.3). However, IPW estimators of HRs are inefficient, yielding wide confidence intervals, especially for minorities and current heavy smokers, due to variability in creating the pseudoweights. CIPW-I, by calibrating the IPW weights to the combined sample of weighted-NHIS and IPW-weighted PLCO, on influence functions of HRs of imputed lung cancer incidence, and poststratifying to lung cancer incidence rates within demographic groups in the population, substantially reduced the variance of the IPW estimators. The CIPW method performed similarly with the CIPW-I method in estimating pure risks (Figures S.2, S.3, Table S.5). This pattern is consistent with the simulation results in Scenarios 1 and 3, because lung cancer is a highly fatal cancer, and gap time between cancer incidence and mortality is not significantly correlated with any of the cancer incidence predictors (Table S.6). For simplicity, the following discussion will focus on CIPW-I only.

Table 4 PLCO Estimates of hazard ratios for lung cancer incidence with 95% confidence intervals using naïve, IPW, CIPW and CIPW-I methods

| Covariates | Naïve | IPW | CIPW | CIPW-I |
|---|---|---|---|---|
| Age | 1.07 (1.06, 1.08) | 1.08 (1.06, 1.10) | 1.06 (1.04, 1.07) | 1.06 (1.05, 1.07) |
| **Race/Ethnicity** (ref: NH White) | | | | |
| NH Black | 1.28 (1.10, 1.50) | 1.27 (0.98, 1.66) | 0.89 (0.82, 0.96) | 0.88 (0.81, 0.95) |
| Hispanic | 0.64 (0.41, 0.99) | 0.34 (0.13, 0.90) | 0.56 (0.51, 0.62) | 0.55 (0.49, 0.61) |
| NH Other | 0.79 (0.6, 1.04) | 0.73 (0.48, 1.11) | 0.81 (0.76, 0.86) | 0.80 (0.75, 0.85) |
| **Sex** (ref: Male) | | | | |
| Female | 0.87 (0.77, 0.97) | 0.83 (0.67, 1.03) | 0.86 (0.81, 0.91) | 0.84 (0.80, 0.89) |
| Education Level | 0.91 (0.89, 0.94) | 0.88 (0.83, 0.93) | 0.92 (0.88, 0.96) | 0.91 (0.87, 0.96) |
| **BMI** (ref: Normal) | | | | |
| Overweight | 0.86 (0.76, 0.97) | 0.82 (0.67, 1.01) | 0.80 (0.69, 0.93) | 0.82 (0.70, 0.95) |
| Obese | 0.80 (0.70, 0.92) | 0.70 (0.54, 0.91) | 0.80 (0.66, 0.97) | 0.80 (0.66, 0.99) |
| Underweight | 1.26 (0.92, 1.74) | 1.10 (0.54, 2.23) | 1.43 (0.89, 2.29) | 1.56 (0.96, 2.53) |
| **Smoking status** (ref: never smoker) | | | | |
| former, quit years>30 | 1.18 (0.86, 1.61) | 1.16 (0.69, 1.96) | 0.89 (0.58, 1.35) | 0.84 (0.55, 1.28) |
| former, 30>quit years>10 | 1.17 (0.83, 1.64) | 0.80 (0.45, 1.41) | 0.79 (0.50, 1.23) | 0.78 (0.49, 1.26) |
| former, quit years<10 | 1.32 (0.82, 2.11) | 0.95 (0.48, 1.88) | 0.83 (0.46, 1.53) | 0.84 (0.45, 1.59) |
| current, cig/day<10 | 4.00 (2.26, 7.07) | 4.13 (1.96, 8.71) | 3.86 (1.96, 7.59) | 4.40 (2.16, 8.98) |
| current, 10<cig/day<20 | 6.81 (3.76, 12.33) | 5.99 (2.60, 13.8) | 5.67 (2.79, 11.54) | 6.34 (3.07, 13.09) |
| current, cig/day>20 | 9.99 (5.52, 18.09) | 9.98 (4.32, 23.07) | 8.92 (4.40, 18.11) | 9.68 (4.66, 20.09) |
| Smoking year | 1.06 (1.05, 1.08) | 1.06 (1.04, 1.08) | 1.06 (1.04, 1.07) | 1.06 (1.04, 1.08) |
| Smoking year×current | 0.97 (0.95, 0.99) | 0.96 (0.95, 0.98) | 0.96 (0.93, 1.00) | 0.96 (0.94, 0.98) |

In addition to improving efficiency of IPW, CIPW-I also substantially changed the HR estimators for older people and minorities. IPW overestimated incidence for people older than 70 (Figure 2), yielding an overestimate of HR for age. For the naïve and IPW-weighted PLCO, Non-Hispanic Blacks had an elevated hazard HR=1.28. However, after calibration (and imputation) on influence functions in the combined PLCO and NHIS sample, and poststratification, the CIPW-I age-specific incidence rates were closer to that in the population (Figure 2) over time. The CIPW-I HR for age dropped to 1.06 and the HR for Non-Hispanic Blacks dropped to 0.89 (Table 4). This happened because PLCO underestimated lung cancer incidence rates for all races except Non-Hispanic Blacks (Table S.4).

Table 5 shows pure risks, Jackknife standard errors, and confidence intervals for 3 non-Hispanic Black individuals with low, medium, and high risks at time $t = 4$ and $t = 7$: (1) a never smoker, (2) a former smoker quitting smoking between 10 to 30 years and having been smoking for 17 years before quitting, and (3) a current smoker consuming 10 to 20 cigarettes per day for 51 years. They were the 10th, 50th, and 95th percentiles of the CIPW-I estimate of $r(t = 7)$ in the CIPW-I weighted PLCO sample, with the respective pure risk estimates being 0.302%, 0.775% and 8.161% at $t = 7$ years. Notice that all the pure risk estimates in Section 5 use the PAR cumulative baseline hazard estimates with the corresponding naïve (weight=1), IPW, CIPW, and CIPW-I pseudoweights.

Table 5 Estimates of pure risks of lung cancer incidence in $t = 4$ and $t = 7$ years for three individuals of low-, medium- and high-risk, with Jackknife variance and 95% confidence intervals using naïve, IPW, CIPW and CIPW-I methods

|  | **Low-Risk ($\times 10^{-4}$)** | | | **Medium-Risk ($\times 10^{-4}$)** | | | **High-Risk ($\times 10^{-4}$)** | | |
| --- | --- | --- | --- | --- | --- | --- | --- | --- | --- |
|  | Est. | JK SD | 95% CI | Est. | JK SD | 95% CI | Est. | JK SD | 95% CI |
| $t = 4$ years | | | | | | | | | |
| Naïve | 1.54 | 0.19 | (1.16, 1.92) | 6.58 | 0.74 | (5.14, 8.02) | 85.74 | 9.44 | (67.25, 104.23) |
| IPW | 1.64 | 0.31 | (1.04, 2.24) | 4.47 | 0.75 | (3.00, 5.94) | 55.95 | 8.99 | (38.33, 73.56) |
| CIPW | 1.48 | 0.18 | (1.12, 1.83) | 3.89 | 0.46 | (2.99, 4.78) | 43.62 | 4.54 | (34.72, 52.53) |
| CIPW-I | 1.55 | 0.19 | (1.17, 1.93) | 3.97 | 0.48 | (3.03, 4.91) | 42.60 | 4.65 | (33.48, 51.72) |
| $t = 7$ years | | | | | | | | | |
| Naïve | 3.01 | 0.38 | (2.28, 3.75) | 12.83 | 1.43 | (10.03, 15.63) | 160.77 | 16.98 | (127.49, 194.05) |
| IPW | 3.23 | 0.61 | (2.04, 4.42) | 8.79 | 1.47 | (5.90, 11.68) | 107.35 | 16.96 | (74.11, 140.58) |
| CIPW | 2.89 | 0.35 | (2.19, 3.58) | 7.59 | 0.89 | (5.86, 9.33) | 83.54 | 8.55 | (66.78, 100.30) |
| CIPW-I | 3.02 | 0.38 | (2.28, 3.76) | 7.75 | 0.93 | (5.93, 9.58) | 81.61 | 8.76 | (64.45, 98.77) |

**Low**-risk: 60-year-old, Non-Hispanic Black female never smoker, some college education, and an overweight BMI.
**Medium**-risk: 63-year-old, Non-Hispanic Black male former smoker quitting 10-30 years, and smoking for 17 years, post-high school training and an overweight BMI.
**high**-risk: 61-year-old, Non-Hispanic Black male current smoker consuming 10-20 cigarettes per day, and smoking for 51 years, <8 years of education, and normal weight.

The naïve PLCO and IPW pure risk estimates for the low-risk individual were close to the CIPW estimates. However, for the medium- and high-risk individuals, the naïve method yielded highest estimates because of the large HRs for age, Non-Hispanic Black, and smoking. For the high-risk individual, the IPW pure risk estimate was higher than the CIPW-I estimate, due to the large IPW estimates of HRs for age and minorities (Table 4). Also, the CIPW-I estimators were much more efficient than the IPW estimators. Pure risk estimates at other time points had the similar pattern (Figure S.3 in Supplementary Materials D).

We evaluated generalizability of the pure risk model by calculating the ratio of expected 7-year lung cancer incidence risk for each method to the observed 7-year lung cancer incidence rate in demographic groups in the US provided by CDC (Table 6). The naïve PLCO and the IPW methods overestimated the risk for high-risk groups, such as older people and minority race groups. The CIPW and the CIPW-I methods generally yielded the closest average risk to the population incidence rates (although not exactly the same) among all methods, especially for minorities.

Table 6 Ratio of the expected and the observed lung cancer incidence risk in the overall PLCO sample and in age, sex, race/ethnicity subgroups using naïve, IPW, CIPW, and CIPW-I methods.

|  | Overall | Age | | | | Sex | | Race/Ethnicity | | | |
| --- | --- | --- | --- | --- | --- | --- | --- | --- | --- | --- | --- |
|  |  | 55-59 | 60-64 | 65-69 | 70-74 | Male | Female | NH-White | NH-Black | Hispanic | NH Other |
| Naive | 1.06 | 1.03 | 1.01 | 1.12 | 1.31 | 1.08 | 1.01 | 0.99 | 1.73 | 1.47 | 1.20 |
| IPW | 1.08 | 0.85 | 0.89 | 1.06 | 1.33 | 1.13 | 1.00 | 1.00 | 1.63 | 0.68 | 1.08 |
| CIPW | 1.03 | 1.07 | 0.96 | 0.99 | 1.11 | 1.04 | 1.02 | 1.03 | 1.05 | 1.04 | 1.01 |
| CIPW-I | 1.03 | 1.07 | 0.96 | 0.99 | 1.11 | 1.04 | 1.02 | 1.03 | 1.05 | 1.04 | 1.01 |

# 6 DISCUSSION

Lung cancer risk estimation is limited by cohorts not representing a target population, and by surveys not typically collecting time-to-disease incidence outcomes. The IPW cohort risk estimators tend to be inefficient, and sensitive to propensity model specification. We take advantage of the fact that health surveys generally have time-to-specific mortality outcomes to propose novel calibrated pseudoweighting methods that improve robustness and efficiency of the IPW estimators by leveraging individual-level cancer-specific mortality outcomes in the survey. The CIPW method creates auxiliary variables using influence functions of the cancer mortality log-HR's estimated from the combined (pseudo)weighted cohort and survey samples. The CIPW-I method first imputes cancer incidence time for cancer specific mortality cases in the survey sample, and then uses influence functions of log-HR's for (imputed) cancer incidence estimated from the combined (pseudo)weighted samples.

Our application for lung cancer incidence risk estimation and simulations show both methods can reduce bias due to misspecified propensity model, and improve efficiency of the IPW estimates of HRs, baseline hazard, and pure risks regardless of cohort self-participation mechanism (noninformative or informative), or the fatalness of the disease. We recommend both CIPW and CIPW-I for lung cancer due to its high fatality. For other less fatal diseases, CIPW-I is recommended when the survival time after diagnosis is not associated with the risk factors of disease incidence because it is more efficient. However, when the survival time after diagnosis is associated with the risk factors of disease incidence, the CIPW-I estimates can be biased. CIPW recommended in this case. In practice, the gap time between cancer incidence and mortality often depends more on overall lethality, the stage of cancer at diagnosis, and tumor markers rather than prediagnostic risk factors (Ewertz et al., 1991), which supports feasibility of the CIPW-I method for cancer risk models.

In the fitted lung cancer pure risk model, PLCO underestimated lung cancer incidence in all demographic groups, except for Non-Hispanic Blacks, underrepresented current smokers, but overrepresented heavy smokers, which implies that the PLCO may overestimate the HRs for Non-Hispanic Blacks and current smoking. The IPW method reduced the HR for current smokers, but yielded inefficient estimates, and did not reduce HR for Non-Hispanic Blacks. In contrast, both CIPW and CIPW-I reduced the naïve HR estimates for Non-Hispanic Blacks and current smokers, and improved efficiency over the IPW estimates.

The final individualized risk model for use in lung cancer screening should involve a model for competing risk of other-cause mortality to estimate the absolute risk of lung cancer incidence (Pfeiffer & Gail, 2017). Several strategies can be employed to obtain a model for competing risk of other-cause mortality. One option is utilizing existing models derived from representative national surveys (Katki et al., 2016). Alternatively, the CIPW or the CIPW-I method can be applied again to develop a new competing risk model from the cohort. Further investigation is necessary to determine the most suitable approach. The resulting methodology will then be applied to develop an individualized lung cancer risk model in the Lung Cancer Cohort Consortium for use in screening.

Our methods develop pure risk models for disease incidence when disease incidence outcomes are never available in a survey, and hence also have some limitations. First, the proposed method can only improve efficiency of the HR estimation for predictors available in both cohort and survey samples. Second, the calibration methods require that the survey sample has enough cases of mortality due to the disease of interest. However, because of small sample sizes, survey samples may have few deaths from common, but usually nonfatal diseases, such as breast cancer, which limits the efficiency and applicability of the proposed calibration methods. Better imputation methods are needed to develop models for diseases that are usually not fatal.